# INFORMATION DIFFUSION IN INTERCONNECTED HETEROGENEOUS NETWORKS

*Shahin Mahdizadehaghdam, Han Wang, Hamid Krim*

North Carolina State University,
Department of Electrical and Computer Engineering,
Raleigh, NC

*Liyi Dai*

Army Research Office,
RTP, Raliegh, NC

## ABSTRACT

In this paper, we are interested in modeling the diffusion of information in a multilayer network using thermodynamic diffusion approach. State of each agent is viewed as a topic mixture represented by a distribution over multiple topics. We have observed and learned diffusion-related thermodynamical patterns in the training data set, and we have used the estimated diffusion structure to predict the future states of the agents. A priori knowledge of a fraction of the state of all agents changes the problem to be a Kalman predictor problem that refines the predicted system state using the error in estimation of the agents. A real world Twitter data set is then used to evaluate and validate our information diffusion model.

## 1. INTRODUCTION

Different topics are discussed/diffused in social networks and because of the distinct dynamics of these diffusions, agents in the social networks are affected in different extents. The temporal and spatial dynamics of diffusion have been studied through sequences of activation nodes and observed as spreading cascades in a network [1].

The information diffusion process models can be classified into three major groups, probabilistic models, thermodynamic models, and counting models. NETINF [2], NETRATE [3] and INFOPATH [4] are the probabilistic models which infer the underlying diffusion network among information sources using consecutive hit times of the nodes by a specific cascade. The main idea behind the thermodynamic models [5, 6, 7, 8] is that heat will propagate from a warmer region to a colder region or gas will move from the region with higher density to the region with lower density. Modeling the information as heat or gas, we can write the rate at which information is changing in agent $i$ as: $\frac{d\psi_i}{dt} = D \sum_j \mathbf{A}(i,j)(\psi_j - \psi_i)$. Where $\psi_i(t)$ is the state of the $i^{th}$ agent at time $t$, $D$ is the diffusion constant which reflects the amount of information passing from an agent to another agent in a small interval of time, and $\mathbf{A}(i,j)$ is the $i,j$ element of adjacency matrix. The counting models [9] form counting processes to find the number of nodes in each group of susceptible or infected nodes. SIS [10] and SIR [11] models are two of the well-known ones in this group.

One central challenge in modeling information diffusion is to understand the structure of the cascades; the existence of unknown external influence factors and unclear graph connections obscures this query. In this paper, we have for the first time modeled the simple diffusion on a general multilayer network and applied the model to publication networks and social media data.

The multilayer network connectivity structure has been proposed and studied in [12, 13, 14, 15, 16], in particular, a two-layer network has been studied in [7] under the Laplacian dynamics. However, the information flow following a Laplacian process on a multilayer network in its most general form has yet to be studied. A multilayer network (illustrated in Fig.1) takes into consideration additional connection possibilities when the true connections for the agents are uncertain. For example, social media such as blogs consist of a set of documents generated by bloggers over time; these documents may share some topical similarities in spite of the different sources they are generated from. It is possible to structure the document similarities into the relational property as a network of documents. Then one can further associate this layer of connection with the connection of bloggers (e.g., according to following-follower connection). This leads to a multilayer network model where information diffuse both within a single layer and across the inter-layers. The additional paths due to the multilayer structure will diffuse the information at a secondary degree, specifically, an absence of a direct connection between two bloggers may be reestablished by counting the topical similarities between the documents they are associated with. We will refer to the resulting network from a multilayer construction as an interconnected network of heterogeneous nodes.

In a community-like network setting with common interest, such as a network of professors with their publications, online forum community, we have observed diffusion related thermodynamical patterns. It turns out that the simple diffusion model on the interconnected network structure is very effective in predicting the future state of the agents.

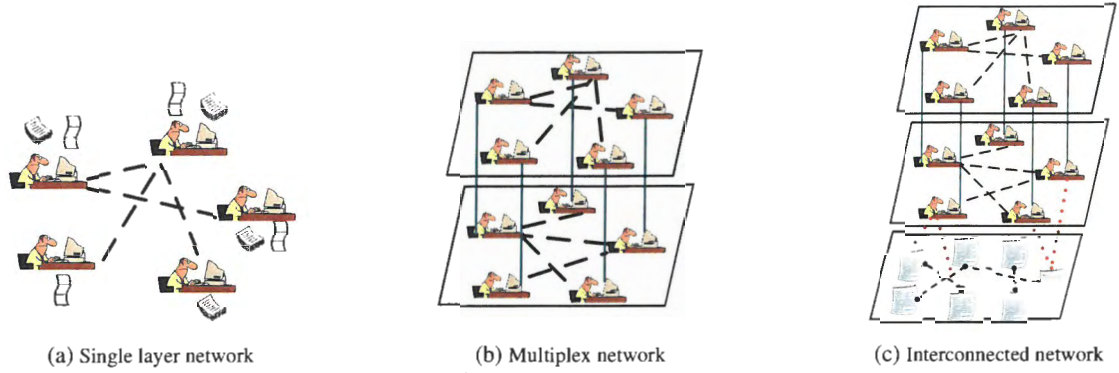

(a) Single layer network  (b) Multiplex network  (c) Interconnected network

**Fig. 1**: a) A single layer. Dashed edges show connections among bloggers. b) A multiplex network. The top layer is based on the hyperlink inter-connectivity of the bloggers, and the bottom layer is the friendship network between the bloggers. The straight inter-layer edges are showing that the bloggers are the same people in both layers. c) An interconnected network of heterogeneous nodes of agents and documents. The dotted edges are showing which blogger (agent) has produced which document.

The paper is organized as follows: We propose our new approach in Section 2, and present substantiating experimental results in Section 3. We finally provide the concluding remarks in Section 4.

## 2. THE PROPOSED METHOD

In closed systems, all changes in the states of agents are a result of interaction of the agents in the network. In a single layer network, a diffusion process based on heat equation has been studied in [5, 6, 7, 8]. Independent of the nodes in a multilayer network being agents or documents, we generalize the single layer diffusion process to:

$$\frac{d\mathbf{X}}{dt} = -\mathcal{L}\mathbf{X}. \qquad (1)$$

Where $\mathbf{X}$ is a $N \times T$ matrix, $N$ being the number of nodes in the multilayer network, and each row of $\mathbf{X}$ is storing a $T$-dimensional state vector for each node. $\mathcal{L}$ is a $N \times N$ supra-Laplacian matrix defined in proposition (1).

**Proposition 1** *We can generally write the supra-Laplacian matrix of an $M$ layer multiplex network with $N$ nodes in each layer as Eqn. (2). Where $\mathcal{L}_L$ is the supra-Laplacian matrix of the intra-layer connectivity and $\mathcal{L}_I$ is the supra-Laplacian matrix of the inter-layer connectivity. $\mathcal{L}_L$ may be in turn, written as direct sum of the Laplacian matrices of the independent intra-layer connectivities:*

$$\mathcal{L} = \mathcal{L}_L + \mathcal{L}_I, \qquad (2)$$

$$\mathcal{L}_L = \bigoplus_{\alpha=1}^{M} D^{(\alpha)} L^{(\alpha)} = \begin{bmatrix} D^{(1)}L^{(1)} & & & \\ & D^{(2)}L^{(2)} & & \\ & & \ddots & \\ & & & D^{(M)}L^{(M)} \end{bmatrix} \qquad (3)$$

*The $D^{(i)}$ is the intra-layer diffusion constant of nodes in layer $i$, and $L^{(M)}$ is the Laplacian matrix of the intra-layer connectivity of layer $M$. The inter-layer supra-Laplacian can be written as $\mathcal{L}_I = \sum_{\alpha=1}^{M}(\mathcal{K}_I^\alpha - \mathcal{W}_I^\alpha)$, where the $\mathcal{K}_I^\alpha$ is the diagonal inter-layer degree matrix of layer $\alpha$, showing the inter-layer degree of the nodes in layer $\alpha$ and the $\mathcal{W}_I^\alpha$ is the inter-layer connectivity matrix of the nodes in layer $\alpha$ with the nodes in the other layers. The $\mathcal{K}_I^\alpha$ and $\mathcal{W}_I^\alpha$ are formally defined in Eqns. (4 and 5) respectively:*

$$\mathcal{K}_I^\alpha = \mathbf{e}_{(\alpha,\alpha)} \otimes \left( \sum_{\beta=1,\beta\neq\alpha}^{M} D^{(\alpha,\beta)} \mathbf{K}^{(\alpha,\beta)} \right), \qquad (4)$$

$$\mathcal{W}_I^\alpha = \sum_{\beta=1,\beta\neq\alpha}^{M} \left( \mathbf{e}_{(\alpha,\beta)} \otimes (D^{(\alpha,\beta)} \mathbf{W}^{(\alpha,\beta)}) \right), \qquad (5)$$

*where $D^{(\alpha,\beta)}$ is the inter-layer diffusion constant of agents from layer $\alpha$ to agents in layer $\beta$, $\mathbf{K}^{(\alpha,\beta)}$ is the diagonal matrix reflecting the degree of each node in the inter-layer connectivity between layer $\alpha$ and layer $\beta$, $\mathbf{W}^{(\alpha,\beta)}$ quantifies the inter-layer connectivity of the layer $\alpha$ nodes to the layer $\beta$ nodes and $\mathbf{e}_{(\alpha,\beta)}$ is an all 0, $M \times M$, matrix with an only 1 element in $(\alpha, \beta)$. $\otimes$ denotes the kronecker product.*

Proof of proposition (1) can be found in the extended version of the paper [17].

Much of the existing work in information diffusion models have a limited scope (of agents, documents, parameters) when predicting the future state of the nodes. More specifically, agent states may be varied by external sources which are not captured in the network, or by some agent actions which may even to some extent, conflict with the model prediction. Considering an interconnected network with supra-Laplacian matrix $\mathcal{L}$, to address this additional auxiliary input, we are further generalizing Eqn. (1) to an open system model as follows:

$$d\mathbf{X}(t) = -\mathcal{L}\mathbf{X}(t)dt + \mathbf{\Sigma}d\mathbf{B}(t). \qquad (6)$$

$\mathbf{B}(t)$ is a $T \times T$ matrix, whose columns are $T$-dimensional vectors with components as independent standard Brownian motions of variances $\sigma_i$ and $\Sigma$ is $N \times T$ matrix, and each row shows the $\sigma_i$ vector for agent $i$. Inspired by the Ornstein-Uhlenbeck (O.U.) process [18], Eqn. (6) describes the velocity of the topical-state of the nodes as a Brownian motion in presence of friction. In other words, to describe the uncertainty due to external effects, we proceed to view the whole system as a massive Brownian particle. The drift term (first term in right-hand side of Eqn. (6)), however, moves the velocity from a martingale state of $\sigma_i d\mathbf{B}(t)$ towards a consensus (captured by the drift term). Solution of the differential equation in Eqn. (6), can be expressed as: given the states of nodes at time $t_0$ we can predict the states at time $t_1$, $t_1 > t_0$ as follows:

$$\widehat{\mathbf{X}}(t_1) = e^{-\mathcal{L}(t_1-t_0)}\mathbf{X}(t_0) + \int_{t_0}^{t_1} e^{\mathcal{L}(s-t_1+t_0)} \Sigma d\mathbf{B}(s). \quad (7)$$

Where $e^{\mathcal{L}(s-t_1+t_0)}$ is a matrix exponential and itself is a $N \times N$ matrix.

Our proposed learning procedure will evaluate the diffusion constants in the supra-Laplacian matrix $\mathcal{L}$ as well as the $\Sigma$ matrix. To that end, we proceed to minimize the Frobenius norm of the difference between $\mathbf{X}$ and it's predict $\widehat{\mathbf{X}}$, resulting from,

$$\underset{\Sigma, D_1,...}{\arg\min} g = ||\mathbf{X}(t_1) - \widehat{\mathbf{X}}(t_1)||_F. \quad (8)$$

Solving this optimization problem helps us decompose the predicted matrix into two main components on the right-hand side of Eqn. (7), the first term representing the interactions in the network, while the second quantifying the uncertainty which results from auxiliary inputs into the system.

### 2.1. Diffusion Network Estimation (Learning the Supra-Laplacian Matrix)

The supra-Laplacian matrix $\mathcal{L}$ which we use in Eqn. (6) for state prediction, is a result of the network connectivity (refer to Eqn. (2)). In practice, hidden connections are pervasive, introducing uncertainty in the prediction, which are causing the information diffusion to require more than the predefined, explicit connections from the network. To that end, consider observations of $\mathbf{X}(t)$ over $t \in [0, t_1]$, denote $\bar{\mathbf{x}}(t) := vec(\mathbf{X}(t))$, the vectorization of $\mathbf{X}(t)$ to obtain a vector differential system in order to learn the supra-Laplacian matrix $\mathcal{L}$ of Eqn. (6):

$$\dot{\bar{\mathbf{x}}}(t) = \Lambda \bar{\mathbf{x}}(t) + \bar{\mathbf{w}}(t) \, , \, 0 \leq t \leq t_1,$$

and we have $\Lambda = \mathbf{I}_T \otimes (-\mathcal{L})$, the Kronecker product of $T$-by-$T$ identity matrix with $(-\mathcal{L})$ and $\bar{\mathbf{w}}(t)$ is the vectorization of $\mathbf{w}(t) = \Sigma \frac{d\mathbf{B}(t)}{dt}$.

We consider the simple cost function $J = \frac{1}{2}\epsilon^T \epsilon$, where $\epsilon = \bar{\mathbf{x}} - \hat{\bar{\mathbf{x}}}$, and hence for the estimation $\hat{\Lambda}$ we have $\dot{\hat{\Lambda}} = \gamma(\bar{\mathbf{x}} - \hat{\bar{\mathbf{x}}})\bar{\mathbf{x}}^T$ (derivative of $J$ with respect to $\hat{x}$), where the estimation $\hat{\bar{\mathbf{x}}}(t+1) = e^{\hat{\Lambda}}\bar{\mathbf{x}}(t)$, and $\gamma > 0$ is chosen appropriately as the scaling gain. In optimization iterations, the estimated value of $\hat{\Lambda}$ at $i^{th}$ iteration is as follows:

$$\hat{\Lambda}_i = \hat{\Lambda}_{i-1} + \dot{\hat{\Lambda}}_{i-1}.$$

We use $\hat{\Lambda}_0 = \mathbf{I}_T \otimes (-\mathcal{L})$, the graph Laplacian of the explicit following-follower network (as initialization). The learned $\Lambda$ may, however, not be exactly structured as $\mathbf{I}_T \otimes (-\mathcal{L})$, due to dependence of topics in the state space, as well as the non-linearity and non-homogeneity of the diffusion. The resulting error $\epsilon$ shall be considered for the estimation of the noise in the Kalman-Bucy filter as discussed next.

### 2.2. A Refined Prediction: Kalman-Bucy Filtering

In prediction applications, the actual states of some of the nodes are sometimes known, and we want to predict those of all remaining nodes. An example of this may be seen in social networks, where state of the hub nodes, such as famous people or users with less restrictive privacy policies are known to the public, and one is interested in predicting the state of other less accessible users. Having partial knowledge of the states of a fraction of the nodes in the network, changes the state prediction problem to a Kalman predictor problem, and helps to refine the predicted states using a Kalman filter. We propose a Kalman-Bucy filter as the optimal linear predictor for our system, and we write the observation equation as $\bar{\mathbf{y}}(t) = (\mathbf{I}_T \otimes \mathbf{H})\bar{\mathbf{x}}(t) + \bar{\mathbf{v}}(t)$, with $\mathbf{H}$ as a diagonal indicator matrix with 1 in all the observed entries, and 0 in all other entries. $\Lambda$ having been learned (see above Section), Kalamn-Bucy equations may be written as:

$$\dot{\bar{\mathbf{x}}}(t) = \hat{\Lambda}\bar{\mathbf{x}}(t) + \bar{\mathbf{w}}(t), \quad (9)$$
$$\bar{\mathbf{y}}(t) = \mathcal{H}\bar{\mathbf{x}}(t) + \bar{\mathbf{v}}(t), \quad (10)$$

where $\mathcal{H} = \mathbf{I}_T \otimes \mathbf{H}$ and the noises $\bar{\mathbf{w}}(t)$ and $\bar{\mathbf{v}}(t)$ are zero-mean white (temporally) processes, i.e, $E(\bar{\mathbf{w}}(t)\bar{\mathbf{w}}(s)^T) = \mathbf{Q}_t \delta(t-s)$, $E(\bar{\mathbf{v}}(t)\bar{\mathbf{v}}(s)^T) = \mathbf{R}_t \delta(t-s)$, $E(\bar{\mathbf{w}}(t)\bar{\mathbf{v}}(s)^T) = 0$. By considering small time intervals on discretization of the linear continues time system ($\delta_t = 1$), one can write the state equation as $\bar{\mathbf{x}}(t+1) = \hat{\mathbf{F}}\bar{\mathbf{x}}(t) + \bar{\mathbf{w}}(t)$, where $\hat{\mathbf{F}} = \mathbf{I} + \hat{\Lambda}$.

Having Eqn. (10) and $\bar{\mathbf{x}}(t+1) = \hat{\mathbf{F}}\bar{\mathbf{x}}(t) + \bar{\mathbf{w}}(t)$ as the state and observation equations respectively, we can predict and refine the predicted states of the nodes using Algorithm (1). The "learning phase" of the Algorithm (1) is estimating the supra-Laplacian matrix $\hat{\Lambda}$ (see above Section). The second phase of the algorithm, is refining the estimated state of the nodes. Note that $\mathbf{R}_t$ is the covariance of the observational error, and $\hat{\bar{\mathbf{x}}}_{t_2|t_1}$ denotes the linear prediction of $\bar{\mathbf{x}}$ at time $t_2$ given observations up to and including time $t_1$. The state covariance $\Pi_t$ satisfies the *Riccati* equation:

$$\dot{\Pi}_t = \hat{\Lambda}\Pi_t + \Pi_t \hat{\Lambda}^T + \mathbf{Q}_t - \mathbf{G}_t \mathbf{R}_t \mathbf{G}_t^T. \quad (11)$$

**Algorithm 1**
**Learning phase:**
1: $\bar{x}(t) \leftarrow vec(\mathbf{X}(t))$
2: $\hat{\mathbf{\Lambda}} \leftarrow \mathbf{I}_T \otimes (-\mathcal{L})$ ▷ Initial state.
3: **repeat:**
4: $\hat{\bar{x}}(t+1) \leftarrow e^{\hat{\mathbf{\Lambda}}}\bar{x}(t)$
5: $\dot{\hat{\mathbf{\Lambda}}} \leftarrow \gamma(\bar{x} - \hat{\bar{x}})\bar{x}^T$
6: $\hat{\mathbf{\Lambda}} \leftarrow \hat{\mathbf{\Lambda}} + \dot{\hat{\mathbf{\Lambda}}}$
7: **until** $\|\bar{x} - \hat{\bar{x}}\|_2 < \eta$. ▷ Convergence criteria.

**Kalman filter prediction on test data:**
1: $\mathbf{R}_{e,t} \leftarrow \mathbf{R}_t + \mathcal{H}\mathbf{\Pi}_{t|t-1}\mathcal{H}^T$ ▷ Updating.
2: $\hat{\bar{x}}_{t|t} \leftarrow \hat{\bar{x}}_{t|t-1} + \mathbf{\Pi}_{t|t-1}\mathcal{H}^T\mathbf{R}_{e,t}^{-1}[\bar{y}_t - \mathcal{H}\hat{\bar{x}}_{t|t-1}]$
3: $\mathbf{\Pi}_{t|t} \leftarrow \mathbf{\Pi}_{t|t-1} - \mathbf{\Pi}_{t|t-1}\mathcal{H}^T\mathbf{R}_{e,t}^{-1}\mathcal{H}\mathbf{\Pi}_{t|t-1}$
4: $\hat{\mathbf{F}} \leftarrow \mathbf{I} + \hat{\mathbf{\Lambda}}$
5: $\hat{\bar{x}}_{t+1|t} \leftarrow \hat{\mathbf{F}}\hat{\bar{x}}_{t|t}$ ▷ Predicting.
6: $\mathbf{\Pi}_{t+1|t} = \hat{\mathbf{F}}\mathbf{\Pi}_{t|t}\hat{\mathbf{F}}^T + \mathbf{Q}_t$

While the $\mathbf{G}_t$ is the Kalman gain $\mathbf{G}_t = \mathbf{\Pi}_t\mathcal{H}^T\mathbf{R}_t^{-1}$. For simplicity, we further assumed that the errors in the state prediction and observation are Gaussian processes.

The designed algorithm shows the discrete time, state update of the Kalman predictor. The estimated states of the available nodes, $\mathcal{H}\hat{\bar{x}}(t)$, are compared with the state of the available nodes, $\bar{y}(t)$, as measurements observed over time, to evaluate the extent of statistical noise and other inaccuracies in predicting phase.

## 3. EXPERIMENTS

Fig. (2a) shows the prediction based on a network of 79 professors. A three-layer network is formed with the first layer showing if the two professors have ever published in the same venue or not. The second layer being the research group membership of these professors and the third layer showing the similarity network between the papers published by these professors (extracted by LDA [19] topic modeling of the abstract of the papers with dimension 10). State of the agents are the 10-dimensional topical vectors which are the mean of the topical representation of the documents produced by the agents. Over time more documents are getting added and the topical state of the agent are changing. The error measure used in all the experiments is the Frobenius norm of the difference of the estimated state of the nodes and ground truth state of the nodes normalized by the Frobenius norm of the ground truth matrix. As may be seen in Fig. (2a), the prediction method based on a three-layer network achieves a lower error than the prediction based on a single-layer network. Note, the single-layer network does not help in predicting the topical states of the agents. The reason is that there are only 79 agents in this experiment and the co-authorship network between the agents is not particularly suited to predict the future

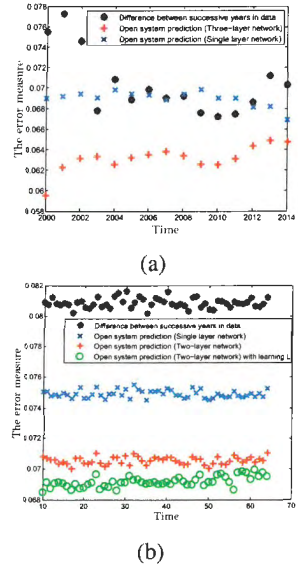

**Fig. 2**: (a) Experiment for college professor network with 1000 publication documents. (b) Experiment over Twitter network with 5000 agents and 8 Hashtags.

state of the agents.

Fig. (2b) is an experiment based on 5000 Twitter users. The first layer is a network among users and the second layer is a network between eight Hashtags used in June 2009. The Hashtags are as follows: #jobs, #spymaster, #neda, #140mafia, #tcot, #musicmonday, #Iranelection, #iremember showing how much these Hashtags are similar to each other by counting and normalizing the number of times they have appeared in the same tweet. The average prediction improvement achieved by the two-layer network is about 13 percent. The prediction by first estimating the Laplacian matrix, has about 15 percent improvement over the single layer prediction method.

Fig. (3) displays the results of an experiment in a small single layer dataset with 300 Twitter agents available. Figs. (3a), (3b), (3c) and (3d) are the same experiments with different observation sizes of 10 percent, 15 percent, 20 percent and 25 percent of state of all the agents in the network respectively. As may be seen in the figures, the prediction error based on the estimated Laplacian matrix yields a lower error than fully trusting the connectivity structure in the network. This as expected, is due to static connectivity network (usually demonstrates the physical or online relation among the agents), falling short on affecting the actual underlying diffusion structure on the network. Having a prior partial knowledge of the agents enabled us to use a Kalman predictor to further refine our prediction.

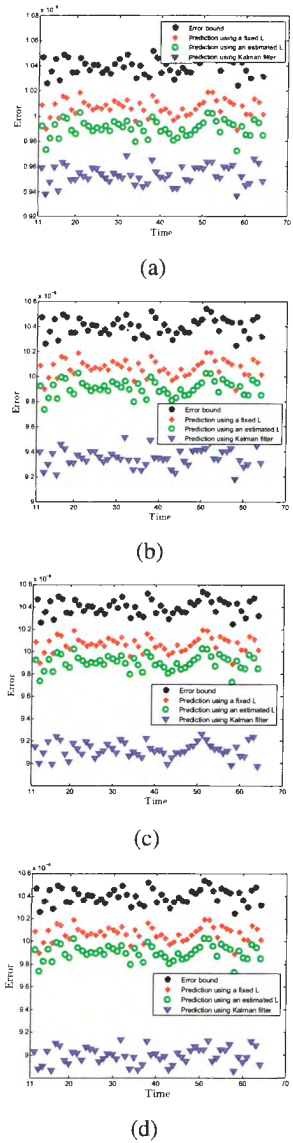

Fig. 3: Experiment over Twitter network with 300 agents. Predicting the state of the agents using a fixed Laplacian matrix, using an estimated Laplacian matrix, and Prediction using Kalman filter with 10 percent, 15 percent, 20 percent, and 25 percent of observation of state of all the agents in Figs. (a), (b), (c) and (d) respectively.

## 4. CONCLUSIONS

In this paper, we proposed a signal processing technique to model and predict states in a multilayer network. The actual diffusion network has been learned by looking at previous diffusion data and has been applied to predicting the future states of agents in the network. Having partial observation of the state of the agents changes the state-space dynamics model to a Kalman filter problem. The Kalman filter allows us to further refine our state prediction by learning over the prediction error of the observed subset. In future work, we will explore the information diffusion problem with a goal to develop necessary tools to analyze diffusion in more complex network structures.